\documentclass[number,times,fleqn]{elsarticle}

\usepackage{graphicx}
\usepackage{amssymb}

\journal{Journal of Computational Physics}

\begin{document}

\begin{frontmatter}
  \title{A partial fraction decomposition of the Fermi function}
  \author{Alexander Croy and Ulf Saalmann}
  \address{Max-Planck-Institute for the Physics of Complex Systems\\
    N\"{o}thnitzer Str.\,38, 01187 Dresden, Germany}
  \begin{abstract}
    A partial fraction decomposition of the Fermi function 
    resulting in a finite sum over simple poles is proposed.
    This allows for efficient calculations involving the Fermi
    function in various contexts of electronic structure or
    electron transport theories.
    The proposed decomposition converges in a well-defined
    region faster than exponential and is thus superior to the
    standard Matsubara expansion. 
  \end{abstract}
  \begin{keyword}
    Fermi function; fractional expansion; numerical methods; 
    electronic structure calculations; electron transport
    theory; finite temperature; Green's function.
  \end{keyword}
\end{frontmatter}

\section{Introduction}
Many problems in electronic structure and electron transport 
calculations involve the evaluation of integrals containing the 
Fermi function. 
These are in general difficult to compute and therefore several
approximation schemes have been developed 
\cite{asme76,ma90,go93,nizh97,ga98,oz07}.
Among them the Sommerfeld expansion \cite{asme76} and the
Matsubara expansion \cite{ma90} being the most prominent ones. 
While the former is by construction useful for low temperatures,
the latter provides in principle a way to cover the range from low to
high temperatures. Moreover, it turns out that the expansion in a (finite) sum 
of simple poles is particularly suitable for evaluating the integrals by making use 
of the residue theorem. For example, finite temperature charge density
calculations only require the evaluation of a Green's function at a finite 
set of energies \cite{wila+95, wawi+99} given by the poles of the expansion. 
Recently the same concept was used for the auxiliary density matrix
propagation in the context of time-resolved electron transport in molecular
wires \cite{wesc+06}. The major disadvantage of the Matsubara expansion
consists in its poor convergence behavior, the error decreasing
only linearly with the number of terms in the expansion.
Here we derive an expansion of the Fermi function in terms of
simple poles with particularly simple coefficients.
We will show that it converges very rapidly with increasing order
of the expansion in a well-defined region which is found to increase
linearly with the order. 

For the following discussion it is convenient to write the Fermi
function $f(\varepsilon)$ in terms of a dimensionless variable $x$,
\begin{equation}
  f(x) = \frac{1}{1+e^x}\quad\mbox{with}\quad x
  = \frac{\varepsilon-\mu}{k T}\;,
  \label{eq:FermiFunc}
\end{equation}
where $\mu$ is the chemical potential, $T$ is the temperature,
$k$ is the Boltzmann factor and $\varepsilon$ denotes the energy.
The expansion consists in finding a partial fraction
decomposition with simple poles of the form
\begin{equation}
  f(x) = \frac{1}{2} - \sum_{p=-\infty}^{\infty} 
  \frac{A_p}{x - x_p}\;,
  \label{eq:FermiExp}
\end{equation}	
where $A_p$ are expansion coefficients and $x_p$ are (possibly
complex) poles.
For practical purposes the sum over $p$ is truncated 
and the Fermi function is approximated by $f(x) \approx f_N(x)$ with
$N$ being the number of terms in the expansion.
	
For example, the well-known Matsubara expansion \cite{ma90} is given in 
terms of the purely imaginary zeros $x_n$ of the denominator in 
Eq.\,(\ref{eq:FermiFunc}), $x_p = \imath \pi (2 p {-} 1)$, which yields 
coefficients $A_p=1$ and gives
\begin{equation}
  f_N(x) = \frac{1}{2} 
  - \sum_{p=1}^{N} \left( \frac{1}{x + \imath \pi (2 p {-} 1)} 
    + \frac{1}{x - \imath \pi (2 p {-} 1)} \right)\;.
  \label{eq:MatsubaraExp}
\end{equation}
For $N\rightarrow\infty$ in Eq.\,(\ref{eq:MatsubaraExp}) the
expansion becomes exact.
However, the convergence is very slow, which renders the
application of this expansion impractical especially for  
low temperatures. 

\section{Partial Fraction Decomposition}
\begin{figure}[h]
  \includegraphics[width=\textwidth]{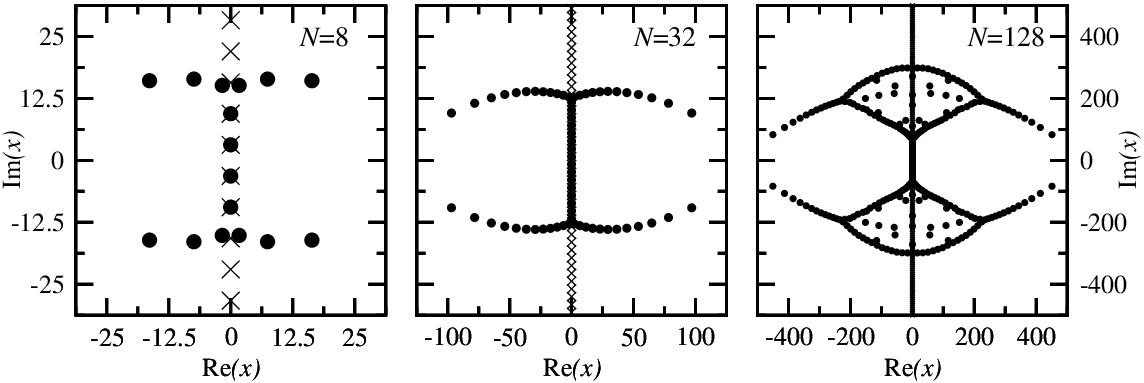}
  \caption{Poles (symbol $\bullet$) of the PFD expansion,
    i.\,e.\ $\pm2\sqrt{z_p}$, 
    with $z_p$ the eigenvalues of matrix (\ref{eq:Zmatrix})
    for various orders $N$.  
    For comparison the  purely imaginary poles (symbol $\times$)
    of the Matsubara expansion (\ref{eq:MatsubaraExp}) are 
    shown as well.
    Note the different scales of the three graphs.} 
  \label{fig:PFDEig}
\end{figure}

The proposed partial fraction decomposition (PFD) is
obtained by firstly writing Eq.\,(\ref{eq:FermiFunc}) as \cite{ga98}
\begin{equation}\label{eq:SinhCosh}
  f(x) = \frac{1}{2} - \frac{1}{2} \tanh(x/2) 
  = \frac{1}{2} - \frac{\sinh(x/2)}{2 \cosh(x/2)} \;,
\end{equation}
and secondly by expanding numerator and denominator in a power
series, truncating the respective sums, such that the degree
of the polynomial in the denominator is larger than the degree
of the numerator polynomial. This procedure gives
\begin{equation}\label{eq:PQ}
  f_{N}(x) = \frac{1}{2} - \frac{1}{2} \frac{P_{N{-}1}(x/2)}{Q_N(x/2)} \;,
\end{equation}	
with polynomials
\begin{equation}\label{eq:PolyQandP}
  P_N(x)=\sum\limits_{m=0}^{N}\frac{x^{2m+1}}{(2m{+}1)!}
  \qquad\mbox{and}\qquad
  Q_N(x)=\sum\limits_{m=0}^N\frac{x^{2m}}{(2m)!}\;.
\end{equation}
This construction allows for a PFD, i.\,e.\ an expansion of the
form 
\begin{equation}
  \frac{P_{N{-}1}(x/2)}{Q_N(x/2)} 
  = \sum_{p=1}^{N} \left(\frac{A_p}{x/2 - x_p}+\frac{B_p}{x/2 + x_p}\right).
  \label{eq:PartFracDecomp}
\end{equation}
Here, $\pm x_p$ are the zeros of the polynomial $Q_N$,
which appear in pairs since $Q_N$ contains only even powers of $x/2$.   
It can be shown that the zeros can be obtained as 
$x_p=\sqrt{z_p}$,
whereby the  $z_p$ are the eigenvalues of the following matrix
\cite{ga98},
\begin{equation}\label{eq:Zmatrix}
Z_{ij} = 2i(2i{-}1)\delta_{j,i+1}-2N(2N{-}1)\delta_{iN}, \quad i,j=1,\ldots,N.
\end{equation}
The eigenvalues can be efficiently calculated using standard methods.
In Fig.\,\ref{fig:PFDEig} we have plotted the poles of
Eq.\,(\ref{eq:PartFracDecomp}), given as
$\pm2x_p{=}\pm2\sqrt{z_p}$, for three sets of eigenvalues $z_p$
for different sizes $N{\times}N$ of the matrix $Z_{ij}$. 
As can be seen from the figure some of the poles $\pm2 x_p$ arising
from the PFD are purely imaginary and are close to the Matsubara
poles. 
On the other hand there are also poles with a
non-vanishing real part, which display an irregular distribution. 
These very poles improve considerably the approximation for the
Fermi function as we show below.

It remains to determine the corresponding expansion coefficients
$A_p$ and $B_p$ in Eq.\,(\ref{eq:PartFracDecomp}).
Multiplying both sides of this equation by $(x/2-x_k)$ and
letting $x\to2x_k$ leaves on the right side of
Eq.\,(\ref{eq:PartFracDecomp}) only the term $A_k$, which is
thus given as 
\begin{equation}
  A_k=\lim\limits_{x\to2x_k}(x/2{-}x_k)\frac{P_{N-1}(x/2)}{Q_N(x/2)}
  =\lim\limits_{\eta\to0}\frac{\eta\:P_{N-1}(x_k{+}\eta)}{Q_N(x_k{+}\eta)}\;.
\end{equation}
By means of the definitions (\ref{eq:PolyQandP}) one finds from this limit
$A_k\equiv1$ and similarly $B_k\equiv1$.
Thus we arrive at the main result of this paper:
The Fermi function can be approximated by the finite sum
\begin{equation}
  f_N(x) = \frac{1}{2} 
  - \sum_{p=1}^{N}\left(\frac{1}{x + 2\sqrt{z_p}}
    +\frac{1}{x - 2\sqrt{z_p}}\right)
  \;,
  \label{eq:PFDExp}
\end{equation}	
with $z_p$ the eigenvalues of matrix (\ref{eq:Zmatrix}).
\begin{figure}[ht]
  \includegraphics[scale=0.55]{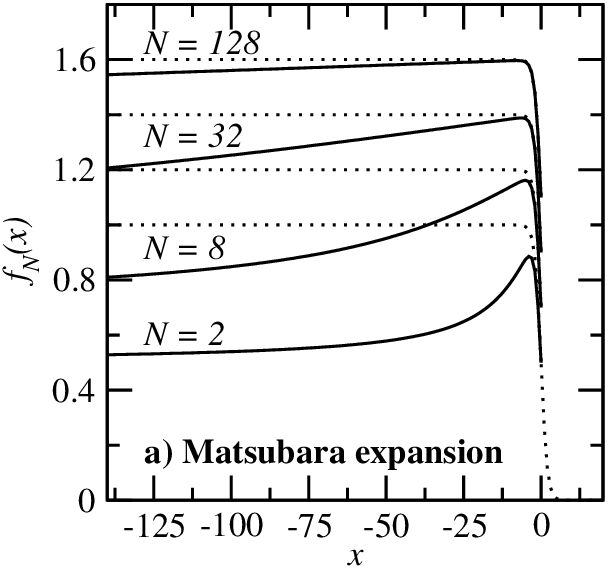}\hfill
  \includegraphics[scale=0.55]{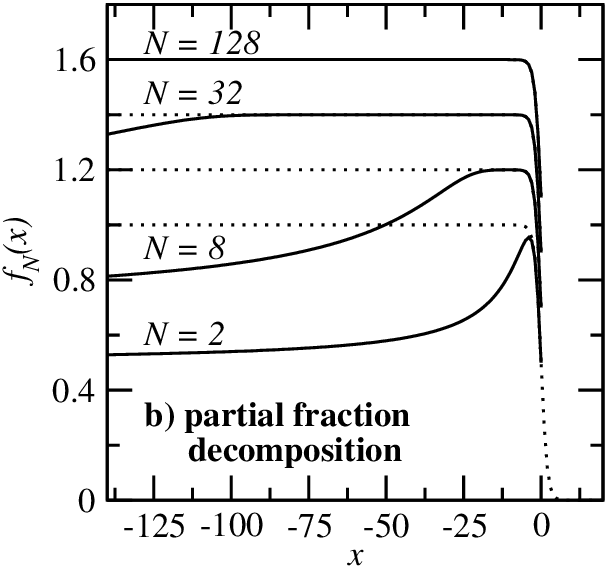}
  \caption{Approximated Fermi function $f_N(x)$
    for expansion orders $N=2, 8, 32, 128$ (full lines).
    Panel a: Matsubara expansion Eq.\,(\ref{eq:MatsubaraExp}),
    panel b: partial fraction decomposition Eq.\,(\ref{eq:PFDExp}).
    The curves are shown for $x<0$ only and are vertically
    shifted by $0.2$ for better visibility. The exact 
	Fermi function Eq.\,(\ref{eq:FermiFunc}) is denoted by dotted lines.
	}\label{fig:Fermi}   
\end{figure}%
The formal structure of this approximation is similar to the 
Matsubara expansion (\ref{eq:MatsubaraExp}).
However, taking advantage of having complex rather than purely 
imaginary poles makes the PFD for given order $N$ of the expansion
vastly superior to the Matsubara expansion. This
can be seen in Fig.\,\ref{fig:Fermi}, where we have shown both expansions 
for different orders $N$.
Whereas the Matsubara expansion (\ref{eq:MatsubaraExp}) in
Fig.\,\ref{fig:Fermi}a does not give a reasonable representation
for any of the orders shown, the PFD expansion (\ref{eq:PFDExp})
in Fig.\,\ref{fig:Fermi}b improves rapidly with increasing order.

\section{Convergence properties}
\begin{figure}[t]
  \includegraphics[scale=0.55]{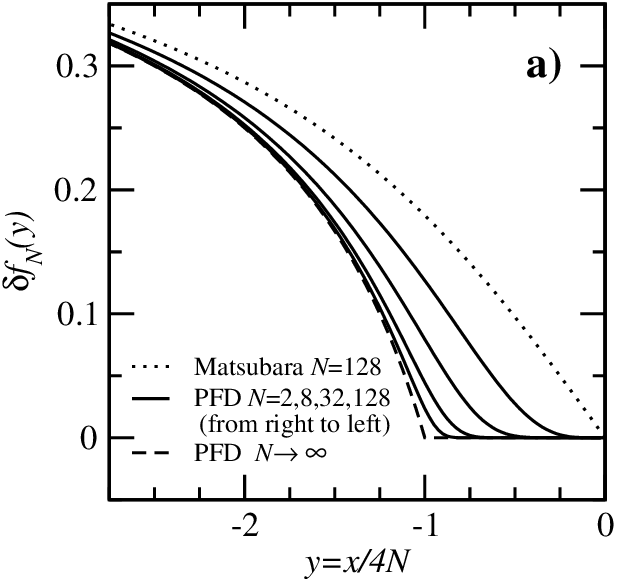}\hfill
  \includegraphics[scale=0.55]{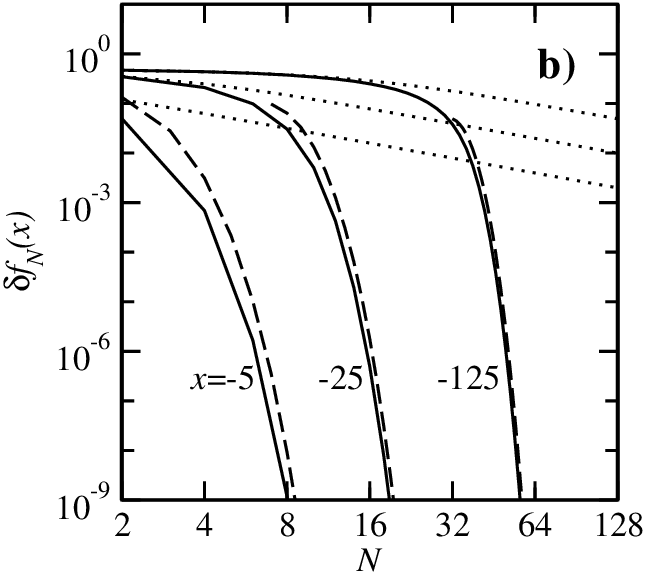}
  \caption{Deviation $\delta f_N$ of the approximated Fermi
    function form the exact one as defined in
    Eq.\,(\ref{eq:DeltaF}). 
    Left panel: $\delta f_N$ as a function of the scaled
    argument $y=x/4N$ for $N=2,\:8,\:32,\:128$ (solid lines)
    and the asymptotic expression (dashed line) as given by
    Eq.\,(\ref{eq:error}). 
    The dotted line shows $\delta f_{128}$ for the Matsubara
    expansion. 
    Right panel:$\delta f_N$ as a function of the expansion
    order $N$ for three arguments $x=-5,\:-25,\:-125$. 
    We compare the Matsubara expansion (dotted lines) and PFD
    (solid lines).
    For the latter case we show also the asymptotic behavior
    according to Eq.\,(\ref{eq:Convergence}) by dashed lines.
    }\label{fig:FermiConvergence}   
\end{figure}

From Fig.\,\ref{fig:Fermi} it becomes clear that the PFD is indeed converging
faster than the Matsubara expansion. In the following we will quantify
the rate of convergence as $N\rightarrow\infty$ and give a range for
$x$ where this convergence behavior can be expected.\footnote{In this section we
  consider only negative arguments $x<0$,  
  but the discussion applies in an analogous manner also to $x>0$.} 
To this end we define the
deviation of the finite expansion from the exact function as
\begin{equation}\label{eq:DeltaF}
  \delta f_N(x) = f(x) - f_N(x).
\end{equation}
Regarding the PFD one makes two observations: First, in terms of the
scaled variable $y=x/4N$ one finds in the limit of large $N$,
\begin{equation}\label{eq:error}
  \lim_{N\to\infty}\delta f_N(x{=}y4N)=
  \left\{\begin{array}{llll}
      0 && \mbox{for} 	& \frac{x}{4N}{=}y\ge-1\\
      \frac{1}{2}\left(1+\frac{4N}{x}\right) =
      \frac{1}{2}\left(1+\frac{1}{y}\right) && \mbox{for}  &  \frac{x}{4N}{=}y\le -1\\
    \end{array}\right.
\end{equation}
The asymptotic function (\ref{eq:error}) is shown along with 
deviations $\delta f_N(y)$ for various finite $N$ in 
Fig.\,\ref{fig:FermiConvergence}a. Second, in the  
range $-4N\le x$, i.\,e.\ $-1<y$, the rate of convergence 
is given by the asymptotic expression
\begin{equation}\label{eq:Convergence}
  \delta f_N(x{=}y4N) \approx  \frac{(x/2)^{2N}}{(2N)!} = \frac{(y2N)^{2N}}{(2N)!},
\end{equation}
which due to the factorial in the denominator decreases faster than
exponential. Eqs.\,(\ref{eq:error}) and (\ref{eq:Convergence}) are the 
main results of this section. They corroborate the statement that the
PFD is expected to yield a better convergence and allow to estimate the
error in actual calculations.

In the remaining part of this section we will justify and discuss 
Eqs.\,(\ref{eq:error}) and (\ref{eq:Convergence}). Considering the
case $y<-1$, one finds from Eq.\,(\ref{eq:PFDExp}), that a finite
expansion behaves as $f_N(x)\propto1/x$ for $x\to-\infty$. Since 
the Fermi function gives $f(x)=1$ for $x\to-\infty$ one expects
qualitatively the behavior given in Eq.\,(\ref{eq:error}). This
holds true for any expansion resulting in a finite sum over
simple poles including the Matsubara expansion, which is shown 
for $N=128$ as dotted line in Fig.\,\ref{fig:FermiConvergence}a. 
In order to verify that this behavior is indeed restricted to $y<-1$,
or equivalently to $x\le-4N$, we write the polynomial $Q_N$ from 
Eq.\,(\ref{eq:PolyQandP}) explicitly as
\begin{equation}\label{eq:QpolY}
  Q_N(x/2)=Q_N(y2N) 
  = \sum\limits_{m=0}^N q_{mN}(y)
  \qquad\mbox{with}\quad
  q_{mN}(y)=\frac{(2N)^{2m}}{(2m)!}y^{2m}.
\end{equation}
Assuming $y<-1$, we see that the ratio of two successive terms
\begin{equation}
  \label{eq:RatioTerms}
  \frac{q_{m\,N}(y)}{q_{m-1\,N}(y)}=y^2 \frac{(2N)^2}{2m(2m{-}1)}
\end{equation}
is always larger than $1$; the terms are monotonically increasing.
Thus terms with $m\gg1$ dominate the sum and we replace the coefficients in $q_{mN}$ 
by the coefficient from $q_{NN}$, 
i.e.\ instead of the sum (\ref{eq:QpolY}) we define
\begin{equation}\label{eq:appQpolY}
  \tilde{Q}_N(y2N)
  = \sum\limits_{m=0}^N \tilde{q}_{mN}(y)
  \qquad\mbox{with}\quad
  \tilde{q}_{mN}(y)=\frac{(2N)^{2N}}{(2N)!}y^{2m}.
\end{equation}
It turns out that in the limit $N\to\infty$ 
this sum becomes equal to $Q_N(y2N)$,  
which can be seen by considering the difference of the newly defined terms in 
Eq.\,(\ref{eq:appQpolY}) from the original terms in Eq.\,(\ref{eq:QpolY}). 
For $m=N-n$ one gets
\begin{equation}
  \label{eq:RelatDeviat}
  1-\frac{q_{N-n,N}(y)}{\tilde{q}_{N-n,N}(y)}
   =1-(2N)^{-2n}\frac{(2N)!}{(2N{-}2n)!}
   =\frac{n/2{-}n^2}{N}+{\cal O}\left(\frac{1}{N^2}\right).
\end{equation}
This expression vanishes for $n=0$ and can be made arbitrarily small by increasing $N$
for all $n\ll N$. 
Terms with larger $n$ can be neglected because they are exponentially small
compared to those with smaller $n$.
Since the sum (\ref{eq:appQpolY}) is a geometric series we obtain
\begin{equation}\label{eq:Qapprox}
  Q_N(y2N)= \tilde{Q}_N(y2N)=\frac{y^2}{y^2{-}1}\:q_{NN}(y)
  \qquad\mbox{for}\quad N\to\infty.
\end{equation}
Analogous considerations for the other polynomial from
Eqs.\,(\ref{eq:PolyQandP}) yield
\begin{equation}\label{eq:Papprox}
  P_{N{-}1}(y2N) \approx
  \frac{(2N)^{2N{-}1}}{(2N{-}1)!}\:\frac{y^{2N+1}}{y^2-1}
  =\frac{y^2}{y^2{-}1}\:p_{N-1,N}(y)
\end{equation}
with $p_{mN}(y)=(y2N)^{2m+1}/(2m{+}1)!$
and we get as an approximation for the ratio, once again using
$N\gg1$, 
\begin{equation}
  \frac{P_{N{-}1}(y2N)}{Q_N(y2N)}\approx\frac{1}{y}=\frac{4N}{x},
\end{equation}
which explains the asymptotic behavior of $\delta f_N(x)$ in
Eq.\,(\ref{eq:error}) for $x<-4N$.

Turning now to the case $y>-1$, we firstly note that there is a
crossover for the ratio (\ref{eq:RatioTerms}) at $|y|N$;
whereas for $m<|y|N$ the terms are increasing, they decrease for
$m>|y|N$.
Thus, for large $N$ the polynomial expression (\ref{eq:PQ})
converges to the exact expression (\ref{eq:SinhCosh}) and the
deviation $\delta f_N(x)$ vanishes as given by
Eq.\,(\ref{eq:error}) for $x>-4N$. 
In order to quantify the rate of convergence it is useful to define
complementary sums to $P_N$ and $Q_N$, namely  
\begin{equation}\label{eq:ComplPolyn}
  \bar{P}_N(x)=\sum\limits_{m=N{+}1}^\infty\frac{x^{2m+1}}{(2m{+}1)!}
  \qquad\mbox{and}\qquad
  \bar{Q}_N(x)=\sum\limits_{m=N{+}1}^\infty\frac{x^{2m}}{(2m)!}\;.
\end{equation}
Therewith the deviation reads
\begin{eqnarray}
  \delta f_N(y4N) &=&  \frac{\sinh(y2N)-\bar{P}_{N{-}1}(y2N)}{2 \cosh(y2N)-2\bar{Q}_N(y2N)}- 
  \frac{\sinh(y2N)}{2 \cosh(y2N)} \nonumber\\
  &\approx& e^{-y2N}\left(\bar{Q}_N(y2N) -\bar{P}_{N{-}1}(y2N)\right)\;.
\end{eqnarray}
The approximation in the second line applies to large values of
$N$. 
We can choose, for any given $y<0$, $N$ sufficiently large such that
$\exp(-y2N){\gg}\exp(+y2N)$. 
The infinite sums defined in (\ref{eq:ComplPolyn}) become small 
compared to the exponentials, $\bar{Q}_N(y2N){\ll}1{\ll}\exp(-y2N)$ and 
$\bar{P}_{N{-}1}(y2N){\ll}1{\ll}\exp(-y2N)$.
It remains to estimate their behavior for large $N$ which can be
done in analogy to the considerations for $Q_N$ and $P_N$,
cf.\ Eqs.\,(\ref{eq:Qapprox}) and (\ref{eq:Papprox}).
Here the ratio of successive terms as defined in
Eq.\,(\ref{eq:RatioTerms}) is always smaller than $1$ and the
first terms in the sum can be used to estimate the sums.
One gets
\begin{equation}\label{eq:PQapprox}
\bar{P}_N(y2N)\approx \frac{1}{1{-}y^2}p_{N,N}(y)
\qquad\mbox{and}\qquad
\bar{Q}_N(y2N)\approx \frac{1}{1{-}y^2}q_{N+1,N}(y).
\end{equation}
This directly leads to Eq.\,(\ref{eq:Convergence}) and concludes the derivation.

Fig.\,\ref{fig:FermiConvergence}b shows this estimate along
with the numerically calculated deviation $\delta f_N$ as a function 
of the expansion order $N$ for selected values of $x$. Even for
small values of $N$ an overall good agreement is found.
Moreover, one sees that the deviation $\delta f_N$ is of order 
$1$ as long as $N <-x/4$. However, for $N>-x/4$ 
(this is where the dashed lines start) it decreases very rapidly 
due to the factorial in the denominator in Eq.\,(\ref{eq:Convergence}).

\section{Conclusions}
We have proposed the expansion (\ref{eq:PFDExp})
of the Fermi function (\ref{eq:FermiFunc}) by using a partial
fraction decomposition. 
Its application requires only the diagonalization of a matrix,
given in Eq.\,(\ref{eq:Zmatrix}),
which has the same dimension $N$ as the expansion.
The expansion converges faster than exponential with
increasing order $N$ for arguments $|x|<4N$.
In other words, the approximation becomes not only more
accurate for higher orders, it can also be used for a wider
range of arguments. An estimate for the error is explicitly given by 
Eq.\,(\ref{eq:Convergence}).
Due to the beneficial convergence properties and the straightforward
implementation we expect the PFD to be of great value in any application based
on an expansion of the Fermi function as sum of over simples
poles. 
Finally, we would like to notice that an analogous expansion can be found for the
Bose-Einstein distribution.

\end{document}